\documentclass[a4paper]{article}

\usepackage{INTERSPEECH2022}
\usepackage{multirow}

\title{Transfer Learning for Robust Low-Resource Children's Speech ASR with Transformers and Source-Filter Warping}
\name{Jenthe Thienpondt, Kris Demuynck}
\address{
    IDLab, Department of Electronics and Information Systems, Ghent University - imec, Belgium
}
\email{jenthe.thienpondt@ugent.be}

\begin{document}

\maketitle
\begin{abstract}
Automatic Speech Recognition (ASR) systems are known to exhibit difficulties when transcribing children's speech. This can mainly be attributed to the absence of large children’s speech corpora to train robust ASR models and the resulting domain mismatch when decoding children’s speech with systems trained on adult data. In this paper, we propose multiple enhancements to alleviate these issues. First, we propose a data augmentation technique based on the source-filter model of speech to close the domain gap between adult and children's speech. This enables us to leverage the data availability of adult speech corpora by making these samples perceptually similar to children's speech. Second, using this augmentation strategy, we apply transfer learning on a Transformer model pre-trained on adult data. This model follows the recently introduced XLS-R architecture, a wav2vec~2.0 model pre-trained on several cross-lingual adult speech corpora to learn general and robust acoustic frame-level representations. Adopting this model for the ASR task using adult data augmented with the proposed source-filter warping strategy and a limited amount of in-domain children’s speech significantly outperforms previous state-of-the-art results on the PF-STAR British English Children’s Speech corpus with a 4.86\% WER on the official test set.
\end{abstract}

\section{Introduction}
Automatic speech recognition (ASR) performance on adult speech data has recently improved noticeably due to the availability of large transcribed speech corpora~\cite{libri, cv_farsi} and the development of end-to-end attention-based acoustic models to leverage the available data~\cite{conformer, rnn_transducer}. However, in low-resource settings, such as children's speech recognition, the performance benefits of these end-to-end models are limited due to the lack of substantial in-domain transcribed data with more traditional approaches such as DNN-HMM ASR models still being competitive~\cite{tlt_2020, slt_2021}.

The recent advances of transfer learning in the field of ASR shows promising results on similar low-resource speech recognition tasks~\cite{wav2vec, wav2vec2, wav2vec_self}. Fine-tuned acoustic frame-level representations from self-supervised models pre-trained with a masking objective on unlabelled adult data can be successfully used for downstream speech recognition applications with a small amount of data~\cite{xls_r}. Encouraging results using pre-trained end-to-end models have recently been established for children's speech recognition with limited amounts of transcribed in-domain data~\cite{tlt_2021, tal_child_2021}. However, these models are pre-trained using large amounts of unlabelled in-domain children's speech, which imposes an important limitation on the usage of this approach. To alleviate this assumption, we attempt to leverage unlabelled adult speech in combination with a data augmentation strategy to tackle the acoustic mismatch between adult and children's speech.

Vocal tract length pertubation (VTLP)~\cite{vtlp} is the most established method to close the domain gap between adult and child speakers~\cite{tlt_2021_vtlp_example}. VTLP applies a linear warping along the frequency axis of the adult speech spectrum to make it perceptually more similar to children's speech. Other work has gained improvements by generating additional input samples by transforming adult speech into children’s speech using a voice conversion model based on cycle GANs~\cite{vc_child_gan, vc_child_gan2}. However, voice conversion models require supplementary child data and introduces an extra training stage which needs to be optimized for the downstream task.

In this work, we propose an augmentation technique based on the source-filter model of speech~\cite{source_filter_model}. We argue that the characteristics of the source and filter component of the speech spectrum behave independently in relation to the adult and children's speech domain mismatch. Subsequently, we introduce a data augmentation strategy which applies a warping function with separate configurations for the source and filter component of the input signal. This enables us to use available adult speech to train more robust acoustic models for transcribing children’s speech. We develop an end-to-end acoustic model based on the recently introduced XLS-R model \cite{xls_r}. This architecture based on wav2vec 2.0 \cite{wav2vec2} is pre-trained self-supervised on large cross-lingual corpora of adult speech with the task of predicting quantized units of masked latent speech representations. Using the proposed source-filter warping strategy enables us to leverage available transcribed adult data and fine-tune the model robustly on the children’s speech recognition task.


\section{Data augmentation}


The disparities between adult and children’s acoustic characteristics poses some inherit challenges to speech recognition systems for children’s speech. Mainly, formants in children’s speech are located at higher frequencies and are prone to higher inter-speaker variability due to the shorter and developing vocal tract of children when compared to adults~\cite{vowel_space_child}. A shorter vocal tract and subsequent higher fundamental frequency ($f_{0}$) also results in undersampling of the spectral envelope due to the widely spaced harmonics~\cite{undersampling_vt, undersampling_vt_recent}. This makes speech recognition methods relying on spectral representations of the input signal less robust when handling children’s speech, especially when trained on adult data, which does not exhibit this problem due to a lower average $f_{0}$. Other distinctions include age-dependent cognitive abilities leading to more frequent disfluencies and mispronunciations~\cite{robust_child_asr}. Several techniques have been proposed to make the spectral representation of adult data more similar to children's speech.

\subsection{Vocal tract length pertubation}
The most used data augmentation strategy to mimic the spectral characteristics of children is applying VTLP~\cite{vtlp} on adult data~\cite{tlt_2021_vtlp_example}. This method applies a linear warping function with a random factor $\eta$ along a range of frequencies covering the significant formants in the spectrum of the signal. In the case of children's speech recognition, $\eta$ is usually constricted to $\eta > 1$ since the average fundamental frequency and formant locations of children's speech is higher in comparison to male and female adult speech.

\subsection{Proposed source-filter warping}
\label{sec:sfw}
In the source-filter model of speech production, speech $y(t)$ is regarded as the convolution of an input signal $s(t)$ and an impulse response $v(t)$, often referred to as the source and filter component, respectively~\cite{source_filter_model}. The resulting equation for speech production in the source-filter model is $y(t) = s(t) * v(t)$. Transferring to the spectral domain, the resulting equation becomes:
\begin{equation}\label{eq:sf_model}
Y(\omega) = S(\omega) V(\omega)
\end{equation}
with the convolution turned into multiplication. $S(\omega)$ and $V(\omega)$ indicate the source and vocal-tract filter spectrum, respectively. VTLP applies a warping function along the frequency dimension contained in $Y(\omega)$, resulting in the usage of the same warping coefficient for both the source and filter component. However, we argue that the optimal warping factor to transform the adult spectrum to a child-like spectral representation is distinct for the source and filter element. Therefore, we propose source-filter warping (SFW), a data augmentation strategy which applies a warping function with separate warping coefficients $\alpha$ and $\beta$ for the source and filter component, respectively.

\subsubsection{Spectral envelope estimation}
\label{sec:spec_env_algo}
We use an iterative smoothing algorithm along the frequency dimension of the power spectrum to estimate the spectral envelope. This reduces the computational complexity as opposed to methods such as cepstral windowing~\cite{cepstral_windowing} and linear predictive coding~(LPC)~\cite{lpc}. Given that $Y_{i}$ represents the power spectrum at frequency location $i$ after applying the short-time Fourier transform (STFT) on the input waveform, the corresponding spectral envelope $V_{i}$ is estimated iteratively with:
\begin{equation}\label{eq:smoothing_func}
V_{i} = max(Y_{i}, V_{i-1} + \gamma (Y_{i} - V_{i-1}))
\end{equation}

with $V_{0} = Y_{0}$ and $\gamma$ being the smoothing factor determining the proclivity of the algorithm to smooth out minor spectral peaks. We apply the algorithm twice: a forward pass starting from the highest frequency bin located at $i_{h}$ and a reverse backward pass starting from the lowest frequency bin. Having estimated the spectral envelope, we can now extract the source component by following $S(\omega) = \frac{Y(\omega)}{V(\omega)}$.

\subsubsection{Warping function}
\begin{figure}[t]
\centering
\includegraphics[scale=0.16]{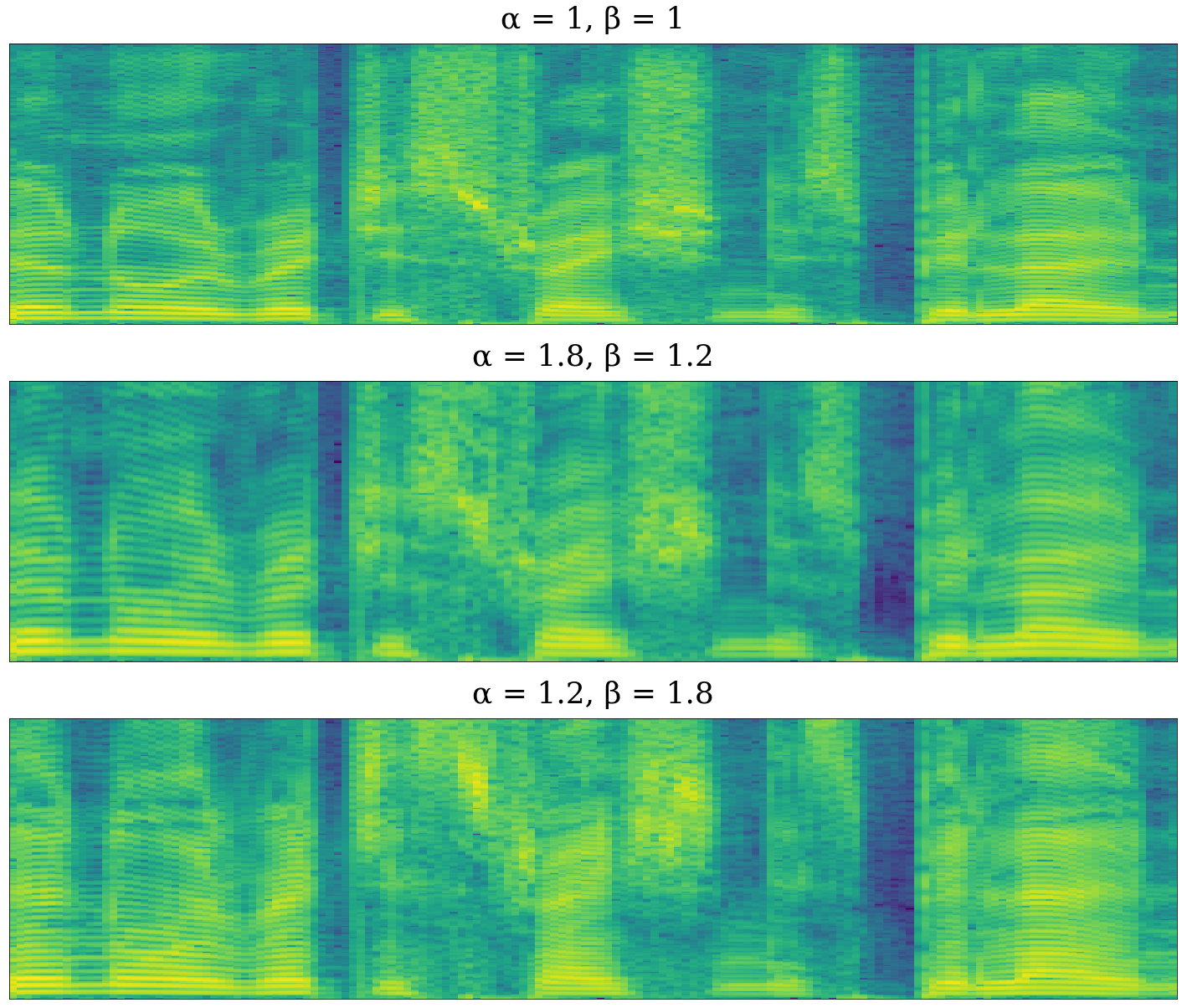}
\caption{Figures of an adult spectrogram (top) augmented by SFW using a high source warping coefficient $\alpha$ (middle) resulting in widely spaced harmonics and a high filter warping coefficient $\beta$ (bottom) which mainly alters the formant locations.} 
\label{fig:warping_example}
\end{figure}

VTLP is typically applied by remapping the center frequencies of the filterbanks in the Mel-spectrogram representation~\cite{vtlp}. However, we do not want to lose spectral resolution inherit due to the subsampling induced by applying Mel-filterbanks on the spectrum. Subsequently, we employ the warping function directly on the extracted source and filter spectrum of the signal. The warped value $F^{\prime}_{i}$ of the source or filter component at frequency bin $i$ is defined as following:
\begin{equation}\label{eq:warping_func}
F^{\prime}_{i} = F_{\lfloor \frac{1}{\lambda} i \rfloor} (1 - (\frac{1}{\lambda} i) \;\mathrm{mod}\; 1) + F_{\lfloor \frac{1}{\lambda} i \rfloor + 1} ((\frac{1}{\lambda} i) \;\mathrm{mod}\; 1)
\end{equation}
with $\lambda$ indicating the warping coefficient. With $\lambda < 1$ and $\lfloor \frac{1}{\lambda} i \rfloor > i_{h}$ for $F_{\lfloor \frac{1}{\lambda} i \rfloor}$ on the right-hand side of Equation \ref{eq:warping_func}, the average power of the 2\% upper frequency bins is used for the resulting $F^{\prime}_{i}$. After applying the warping functions, we can reconstruct the augmented spectrogram by multiplying the source and filter component. Figure \ref{fig:warping_example} illustrates the effect of source-filter warping with varying values for the source warping coefficient~$\alpha$ and filter warping coefficient~$\beta$.
\section{Acoustic modelling}
Traditional ASR systems are based on the DNN-HMM model \cite{dnn_hmm_overview, chain_model_kaldi}. The neural network serves as the acoustic model and estimates the posterior probabilities of the acoustic units in the framed speech signal and is usually implemented as a CNN or TDNN. DNN-HMM models are still popular in low-resource ASR conditions such as children's speech recognition \cite{tlt_2020, tlt_2021}.

\subsection{End-to-end ASR systems}
Recently, attention-based end-to-end speech recognition systems with an encoder-decoder architecture have gained state-of-the-art results on adult ASR benchmarks~\cite{conformer, rnn_transducer}. However, training randomly initialized end-to-end models is known to require a significant amount of training data to model the latent acoustic representations robustly~\cite{deep_speech}. This poses an important limitation on the direct usage of these models in children's ASR applications.

In the context of low-resource speech recognition, promising results are recently gained by applying transfer learning techniques to adapt a model trained on large speech corpora to perform robust speech recognition on resource constrained out-of-domain data~\cite{wav2vec, wav2vec2}. We apply transfer learning on a Transformer model pre-trained on adult speech data with a masking objective to build a robust children's speech ASR system.

\begin{figure}[t]
\centering
\includegraphics[scale=0.18]{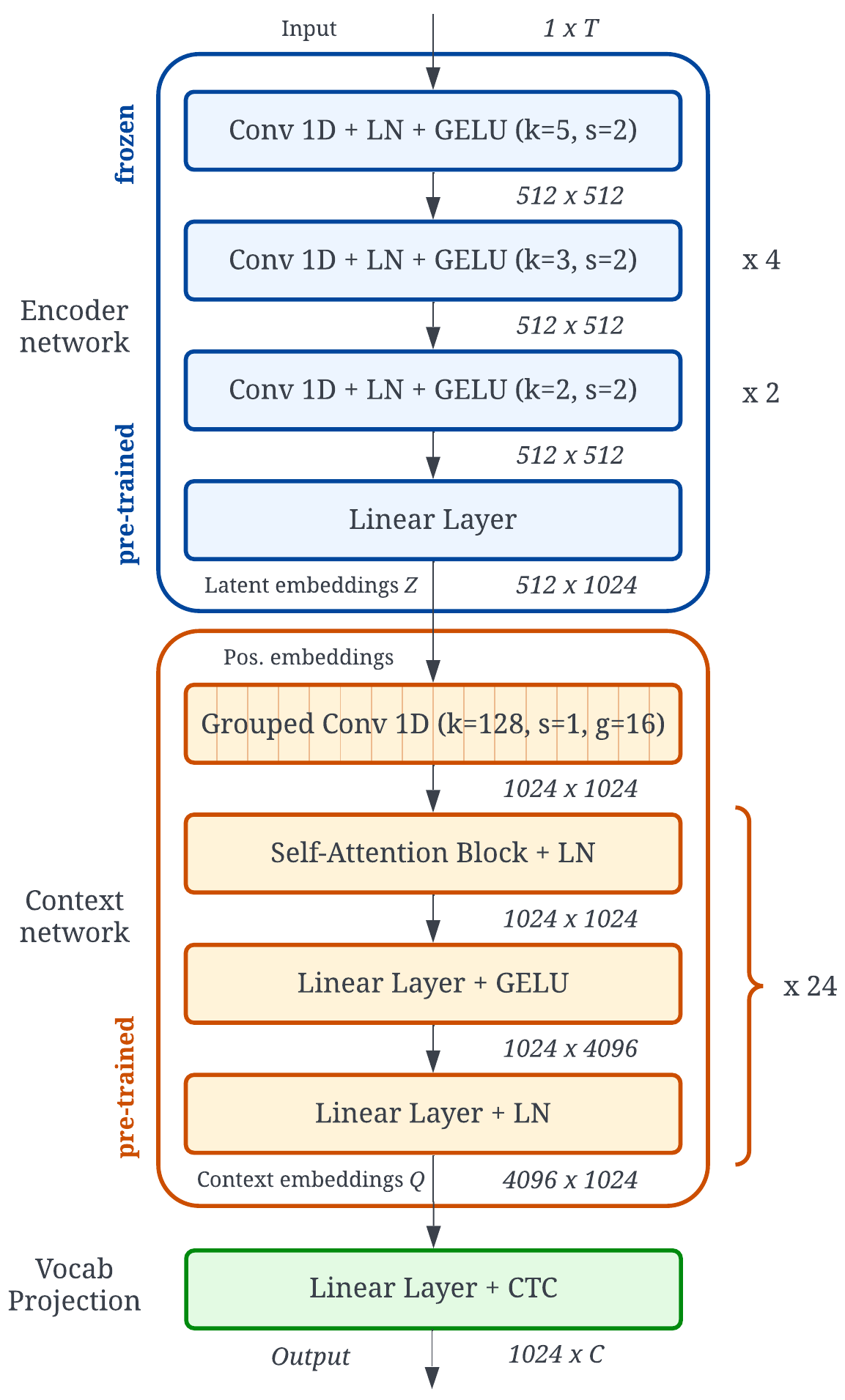}
\caption{Diagram depicting the network architecture for our ASR system. We denote k, s and g for kernel size, stride and group size in the convolutional layers, respectively. T refers to the temporal dimension of the input waveform and C denotes the output dimension matching the vocabulary size.}
\label{fig:network}
\end{figure}

\subsection{Proposed Transformer ASR system}
The architecture of our ASR system is based on the recently proposed XLS-R model~\cite{xls_r}, a Transformer architecture based on wav2vec 2.0 \cite{wav2vec2}. The model consists of two components: the \textit{encoder network} and the \textit{context network}. The encoder network consists of stacked blocks of temporal convolutions followed by layer normalization (LN)~\cite{layer_norm} and the GELU~\cite{gelu} activation function, converting the raw input waveform into a sequence of latent speech representations $Z$. Positional embeddings are added to the speech representations for the context network to be able to model long-range dependencies of the input signal. The positional embeddings are learned by a grouped convolution, providing the following layers relative positional information~\cite{conv_pos_embeddings}. The speech representations are then used as input to the context network. The context network consists of stacked Transformer blocks, modelling contextualized acoustic representations $Q$.

Following~\cite{xls_r}, the model is pre-trained on cross-lingual unlabelled adult speech corpora in a self-supervised manner using a contrastive loss function in which the model needs to predict quantized audio representations from masked output latent representations of the encoder network. The learned context representations should be able to capture robust acoustic and linguistic characteristics of the input utterance. We choose to pre-train on cross-lingual adult data to make the context representations and subsequent ASR system robust against children's speech in a variety of languages without the need to pre-train and optimize separate models in future work.

After pre-training, a linear layer is added to the context network which projects the context representations $Q$ to the vocabulary of the ASR task. We fine-tune the model using a combination of augmented adult speech with the proposed SFW strategy described in Section \ref{sec:sfw} and in-domain children's speech. By fine-tuning the pre-trained adult model on children's speech and augmented adult data, the network should be able to make the learned adult acoustic representations of the context network robust against the corresponding children's speech. The optimization of the model on the ASR task is done using the Connectionist Temporal Classification (CTC)~\cite{ctc_loss} objective function. During this fine-tuning stage, we freeze the layers of the encoder network to prevent overfitting and reduce computational complexity. The final architecture is shown in Figure \ref{fig:network}.

\section{Experimental evaluation}
To analyse the impact of the proposed SFW and transfer learning strategy for children's speech ASR, we evaluate our approach on the test set of the \textit{PF-STAR British English Children’s Speech} corpus \cite{pf_star}. The dataset contains 7.4 and 5.8 hours of transcribed audio for the training and test partitions, respectively. The age of the children in the dataset range from 4 to 14 years. Following other papers \cite{pf_star_2016,vc_child_gan, kathania_pf_star_adult_only}, we use the training subset of the WSJCAM0 corpus \cite{wsjcam0} as out-of-domain adult data, containing 15.5 hours of British English speech across 92 speakers.

\subsection{Source-filter warping}
To apply our proposed SFW augmentation strategy, power spectrograms with a window size of 25 ms and hop length of 10 ms are generated from the adult speech waveforms using an FFT length of 512. Subsequently, we use the algorithm described in Section \ref{sec:spec_env_algo} with $\gamma = 0.2$ to estimate the filter and source component on which we apply our warping function given by Equation \ref{eq:warping_func}. As we warp both components independently, we allow warping coefficients $\alpha$ and $\beta$ to be relatively high by randomly choosing a value between 1 and 1.3. 

Since our end-to-end ASR system acts on the waveform of the input signal, we need to convert the spectral representation back to the temporal domain. The Griffin-Lim algorithm~\cite{griffin_lim, griffin_lim_fast} is used to estimate the phase component of the STFT power spectrograms. To limit the computational impact, the alternating forward and inverse STFT step in the algorithm is only repeated 8 times.

\subsection{Acoustic model training}
Our model is pre-trained self-supervised on unlabelled adult speech data to learn meaningful acoustic representations of speech using the architecture depicted in Figure \ref{fig:network}. More details about the pre-training step are described in~\cite{xls_r}.


During fine-tuning, we pool the training data of the PF-STAR and WSJCAM0 datasets together with equal sampling probability for both domains during batch construction. A batch size of 48 is used and the model is trained for 60K steps using the AdamW~\cite{adamw} optimizer. The learning rate starts at 5e-5, followed by a warmup stage of 500 steps to 1e-4 and then linearly decreases to 0. The input waveform is mean and variance normalized and randomly cropped according to a random start and end timestamp of the transcriptions. The crop size is limited to contain between 2 and 4 seconds of audio with no usage of SpecAugment~\cite{specaugment}. We found this to be more effective and faster to train as opposed to using longer utterances with SpecAugment enabled. A bi-gram in-domain language model (LM) was employed to decode the test utterances, similar to~\cite{vc_child_gan}. The out-of-vocabulary (OOV) rate of the LM is 2.27\% with a perplexity of 70.3 with respect to the PF-STAR test set.

\section{Results}
Table \ref{tab:overall_performance} shows the performance of the proposed transfer learning approach and SFW augmentation strategy. The baseline performance of our ASR system using a language model and no data augmentation strategy trained on children's data gathers a strong baseline result of 6.89\% WER on the PF-STAR test set. Including the out-of-domain adult WSJCAM0 dataset improves performance on the children's test set only negligibly. We suspect this is mainly due to the extensive self-supervised pre-training of the model on adult data. However, employing the SFW strategy during training on the adult dataset, we see a relative improvement of the WER on the children's test set of 28.4\% over the system without adult data augmentation. This shows that the proposed SFW strategy successfully induces the characteristics of children's speech into the adult utterances, closing the domain gap between the adult and children's speech datasets significantly. To the best of our knowledge, this is the best published result on the PF-STAR test set.

\begin{table}[h]
  \caption{WER performance on the PF-STAR test set.}
  \label{tab:overall_performance}
  \centering
  \begin{tabular}{l c c}
    \toprule
    
    \multicolumn{1}{c}{\textbf{Training data}} &
    \multicolumn{2}{c}{\textbf{WER(\%)}} \\
    
    \cmidrule(lr){2-3}
    \multicolumn{1}{c}{} & 
    \multicolumn{1}{c}{\textbf{CTC}} &
    \multicolumn{1}{c}{\textbf{with LM}} \\
    
    \midrule
    WSJCAM0                    & 40.94  & 18.64   \\ 
    PF-STAR                    & 9.53  & 6.89   \\ 
    PF-STAR + WSJCAM0          & 9.48  & 6.79    \\ 
    PF-STAR + WSJCAM0 (SFW)    & \textbf{6.57} & \textbf{4.86}    \\ 
    \bottomrule
  \end{tabular}
\end{table}


A performance analysis of the proposed SFW in comparison to VTLP can be found in Table \ref{tab:sfw_performance}. The baseline system is trained on adult and children's speech without any augmentation strategy. Notably, the artefacts introduced by the temporal reconstruction of the input signal from the FFT spectrum with the Griffin-Lim (GL) algorithm has a beneficial robustness effect, as shown by the \textit{GL} experiment where we did not apply any warping augmentation but did convert the input utterances between the spectral and time domain. 

The best performing VTLP configuration with a random warping factor between 1 and 1.2 improves the WER with 17.8\% relative over the baseline with usage of a language model. The best SFW strategy is gained with $\alpha$ and $\beta$ randomly varying independently between 1 and 1.3 and improves the result further with 12.9\% WER relatively over the best VTLP configuration. Due to the independent warping of the source and filter component, the best SFW configuration allows for higher maximum warping coefficients as opposed to VTLP, which only has one parameter to control the warping.

\begin{figure}[t]
\centering
\includegraphics[scale=0.30]{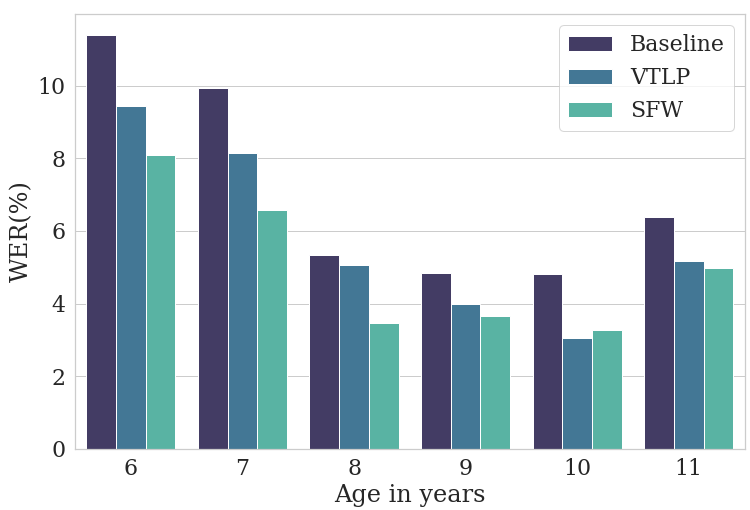}
\caption{Bar chart showing the WER on the PF-STAR test set for each age group using VTLP and the proposed SFW.}
\label{fig:bar_age}
\end{figure}

\begin{table}[h]
  \caption{Performance analysis of source-filter warping.}
  \label{tab:sfw_performance}
  \centering
  \begin{tabular}{c c c c}
    \toprule
    
    \multicolumn{1}{c}{\textbf{Method}} &
    \multicolumn{1}{c}{\textbf{Warp Factors}} &
    \multicolumn{2}{c}{\textbf{WER(\%)}} \\
    
    \cmidrule(lr){3-4}
    \multicolumn{2}{c}{} & 
    \multicolumn{1}{c}{\textbf{CTC}} &
    \multicolumn{1}{c}{\textbf{with LM}} \\
    
    \midrule
    baseline & / & 9.48 & 6.79   \\ 
    \midrule
    \midrule
    GL & / & 8.38 & 6.27 \\ 
    \midrule
    \multirow{4}*{VTLP} & $\eta = [1,1.15]$ & 7.79 & 5.62   \\ 
     & $\eta = [1,1.20]$ & 7.75 & 5.58   \\ 
     & $\eta = [1,1.25]$ & 8.07 & 5.96   \\ 
     & $\eta = [1,1.30]$ & 8.32 & 6.22 \\ 
    \midrule
    \multirow{4}*{SFW} & $\alpha, \beta = [1,1.15]$ & 7.19 & 5.32 \\ 
     & $\alpha, \beta = [1,1.20]$ & 6.92 & 5.07 \\ 
     & $\alpha, \beta = [1,1.25]$ & 6.69 & 4.95 \\ 
     & $\alpha, \beta = [1,1.30]$ & \textbf{6.57} & \textbf{4.86} \\ 
    \bottomrule
  \end{tabular}
\end{table}

Figure \ref{fig:bar_age} shows the WER of each age group in the PF-STAR test set from the baseline system together with the best performing VTLP and SFW configuration from Table \ref{tab:sfw_performance}. We see that the performance increase relative to the baseline model is consistent across all ages, indicating that the proposed SFW is able to model characteristics of varying age groups. Transcription quality is clearly correlated to the age group with the WER being inversely proportional to the speaker age. An exception is the group of age 11, we suspect the degradation is mainly due to the more complex text transcriptions appearing in this age group~\cite{pf_star}. Interestingly, we see that the largest performance improvement of SFW over VTLP is manifested in the younger age groups. We believe this is due to the benefit of independently warping the source and filter component in SFW, as younger children exhibit more varying formant locations and fundamental frequencies as opposed to older children.

\section{Conclusion}
In this paper, we applied transfer learning on a Transformer model pre-trained with adult speech and proposed the source-filter warping data augmentation strategy for robust children's speech ASR. Using a few hours of in-domain children's speech data, our fine-tuned Transformer model scores a WER of 6.79\% on the PF-STAR children's speech test set. Applying our proposed source-filter warping strategy to close the adult and children's speech domain gap improves this strong baseline system with a final WER of 4.86\%, significantly outperforming previous state-of-the-art results on the PF-STAR test set.

\bibliographystyle{IEEEtran}

\bibliography{mybib}

\end{document}